\input{aipcheck.tex}

\documentclass[final]{aipproc}

\layoutstyle{8x11double}

\SetInternalRegister\hbadness{8000} 

\newcommand{\nh}{\thinspace 10^{21} {\rm cm}^{-2}}
\newcommand{\lsim}{\mathrel{\hbox{\rlap{\lower.55ex \hbox {$\sim$}}\kern-.0em\raise.4ex \hbox{$<$}}}}
\newcommand{\gsim}{\mathrel{\hbox{\rlap{\lower.55ex \hbox {$\sim$}}\kern-.0em\raise.4ex \hbox{$>$}}}}
\newcommand{\etal}{{\it et al.}}

\newcommand{\Ms}{M$_\odot$}

\newcommand{\rcm}{\rm cm}
\newcommand{\dens}{\thinspace \rcm^{-3}}

\begin{document}

\title
      []
      {X-Ray Spectroscopy of Gamma-Ray Bursts}

\classification{} \keywords{Gamma-Ray Bursts}

\author{L. Piro}{
  address={Istituto Astrofisica Spaziale Fisica Cosmica, CNR, Roma, Italy},
  email={piro@ias.rm.cnr.it},
  thanks={}
}

\copyrightyear  {2002}

\begin{abstract}
Observational evidence  of iron absorption and emission lines in
X-ray spectra of Gamma-Ray Bursts is quite compelling. I will
briefly review the results, summarize different models and
describe the connection with massive progenitors in star-forming
regions implied by these results. This link is also supported by
measurements of the X-ray absorbing gas in several GRB's, with
column density consistent with that of Giant Molecular Clouds
harbouring star-formation in our Galaxy, as well as by evidences
gathered in other wavelengths. However, the volume density
inferred by the fireball-jet model is much lower than typical of a
GMC, and I will confront this with the alternative explanation of
fireball expansion in a high dense medium, outlining the problems
that both models have at present. Finally I will briefly summarize
some results on dark GRB's, and describe the prospects of high
resolution X-ray spectroscopy in getting closer to the central
environment of GRB,  and far in the Early Universe by using GRB as
beacons to probe star and galaxy formation.
\end{abstract}

\date{\today}

\maketitle

\section{Introduction}

X-ray observations of GRB are playing a key role in several areas
of the GRB research. Most of the fast and precise localizations of
GRB have been obtained in X-rays by BeppoSAX, HETE-2 and Rossi XTE
by combining a gamma-ray burst monitor trigger with the location
accuracy provided by X-ray detectors.

X-ray spectroscopy of the prompt and afterglow emission is
providing important clues as to the origin of GRB progenitors, the
emission processes and the environment of different classes of
GRB. The presence of emission and absorption Fe X-ray lines in
some GRB is indicating massive progenitors in star forming
regions. The gas in the local environment of the GRB or in the
host galaxy affects also the X-ray spectral shape. In this regard,
X-ray spectral measurements can provide information on the column
density of the gas, its chemical composition and ionization stage.

The characterization of the environment is of particular
importance to understand the origin of different classes of GRB's:
optically bright and  dark GRB's, X-ray rich GRB, and short GRB's.
Some of the differences amongst two or more classes could be in
fact the result of a different environment, rather that reflecting
intrinsic differences in the central source. While this could not
be the case of short GRB (thought to be produced by binary mergers
rather than  by massive progenitors
\citep[e.g.][]{w01})\footnote{In this case  the environment would
make a difference in the {\it afterglow} properties because,
lacking an external medium to make an external shock, the
afterglow would be much fainter than in the case of long bursts},
 this sort of unification scenario of GRB could in principle
account for  other classes properties. For example, dark events
could be associated to GRB whose optical light is effectively
estinguished by dust either in the local environment or in the
host galaxy \citep[e.g.][]{ry01}. X-ray rich events could  be
associated with "dirty" fireballs \citep{dermer_dirty99}, i.e.
fireballs that, due to high baryon loading,  expand with a much
lower Lorentz factor, boosting the photons in the X-ray band
rather than in the $\gamma$-ray range.

 Another
important piece of information that could play an important role
in making up different observed properties is the distance. For
example the properties of X-ray flashes and some dark GRB's can be
explained if they lie at $z>5-10$.

In this paper I will review the present picture and discuss future
perspectives in assembling some of the missing elements,
outlining, in particular, the role of X-ray spectroscopy.


\begin{figure}
\centering
\includegraphics[width=1.0\textwidth]{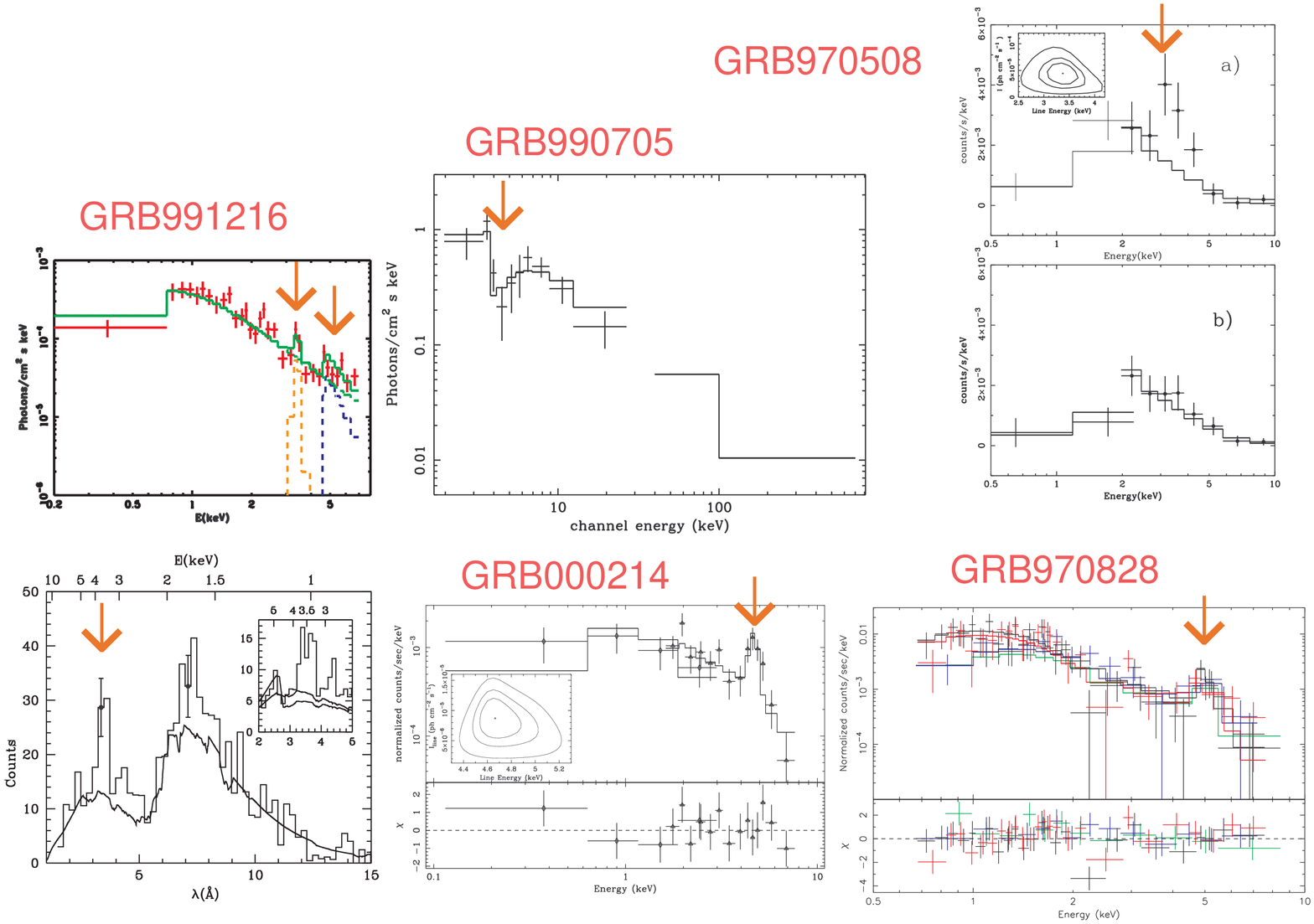}
\caption{Iron features in GRB: GRB970508 \citep{paa+98};
GRB970828: \citep{yym+01}; GRB990214 \citep{apv+00}; GRB991216
\citep{pgg+00});GRB990705 \citep{afv+00})} \label{fig:lines}
\end{figure}


\section{X-ray absorption and emission features}

Iron features are ubiquitous in several classes of X-ray sources
and have been used to probe the emission mechanisms and the close
environment of these sources\citep{piro93}. The observational
evidence of iron features in the X-ray spectra of GRB is quite
compelling. So far, there have been 5 detections of iron features
in GRB (Fig.\ref{fig:lines}), four in the afterglow phase
(GRB970508 \citep{paa+98}; GRB970828: \citep{yym+01}; GRB990214
\citep{apv+00}; GRB991216 \citep{pgg+00}), and one during the
prompt phase (GRB990705 \citep{afv+00}),  each at about 3 sigma
level (with the exception of GRB991216, with a significance above
4 sigma).  In three cases (GRB970508, GRB970828, GRB991216), the
X-ray redshifts derived from the iron lines were consistent with
those from optical spectroscopy. For the dark GRB000214, no
optical redshift is available as yet. Finally, the case of
GRB990705 is particularly important. No optical spectroscopy was
available when the original paper was published\citep{pgg+00}. The
energy of the iron feature implied an X-ray redshift
$z=0.86\pm0.14$. This prediction is now confirmed by optical
spectroscopy, that set the redshift of the host galaxy at
$z=0.8435$ \citep{andersen_990705}. These findings demonstrate
that the measurement of the redshift by X-ray spectroscopy is more
than a mere possibility, and that it can provide reliable results,
a fact that is particularly important when optical spectroscopy is
difficult or not possible at all.

Let us now summarize the observational picture and the ensuing
theoretical implications. For the features in the three afterglows
with an independent optical redshift, we can derive the line
energy at the burst site and therefore the ionization stage of the
medium. FeI to Fe XVII ions  are characterized by a $K_{\alpha}$
line energy at $\approx 6.4$ keV , that rises to  6.7 keV for
He-like ions and to 6.9 keV for H-like ions. The iron edge is
located at 7.1 keV for neutral iron, rising slowly to 8.8 keV for
He-like ions, and then to 9.3 keV for H-like ions.

In GRB991216 (z=1.02) the rest-frame line energy  is
$6.95\pm0.15$. There is also evidence of an additional emission
feature at $9\pm1$keV, that  is associated with a narrow H-like
iron recombination edge {\it in emission} (also known as Radiative
Recombination Continuum, RRC). In the case of GRB970828 (z=0.96),
the emission feature  at $9.3\pm0.5$ has  been also attributed to
the H-like recombination edge in emission. In the case of
GRB970508, the line energy  is $6.25\pm0.55$, i.e. consistent with
an iron stage from neutral to H-like.

There is also marginal evidence of line variability on time scales
of $\lsim day$ in three afterglows (GRB970508, GRB970828 and
GRB000214), while for GRB991216 the observation was too short (10
ksec). In the latter case a line broadening of $\approx 10\%$ is
measured.

The first conclusion that can be drawn is that the line-emitting
gas, {\it in the afterglow phase}, is highly ionized. The most
obvious source of ionization is the GRB itself, that produces a
copious flux of hard X-ray photons ({\it photoionization
scenario}). In this setting, the  medium has to be located
 outside the fireball region (i.e. at $R\gsim 10^{15}$cm:
{\it distant reprocessor
scenario}\citep[e.g.][]{lcg_line99,pgg+00,weth+00}). The line
variability is naturally expected from the light travel time
between the GRB and the reprocessor.

In the early phase (i.e. on a time scale of $\approx 10$s), when
the ionization front is still expanding, a substantial fraction of
the medium in the line of sight is still to be ionized, thus
producing X-ray absorption features, that will disappear when the
medium is becomes completely ionized \citep[e.g.][]{pl98,bdcl99}.
A transient iron absorption edge is then expected if the medium
lies in the line of sight, exactly what has been observed in
GRB990705. On longer time scales, ($\approx t_{rec}$, the
recombination time scale), electrons recombine with ions,
producing the line and the RRC \citep[e.g.][]{pkhl00,pgg+00},
observed in the afterglow phase. This scenario requires the
presence of a dense medium with high iron overabundance ejected
before GRB , with a velocity (implied by the line width) of
$\approx 0.1c$ and a mass of iron $\gsim 0.01$ \Ms
\citep[e.g.][]{pgg+00}. The most straightforward scenario that
emerges is that of a massive progenitor that ejects, before the
GRB, a substantial fraction of its mass in a SN-like explosion, as
in the case of the SupraNova model \citep{vs_supra99}.

In the ({\it nearby reprocessor scenario}
\citep{rm_line00,mr_line01}) ionizing photons would be produced by
post-burst activity of the "remnant" of the central source after
the GRB event.  In this case the reprocessor, associated with the
outer stellar envelopes of a massive progenitor, can be much
nearer to the central source. Line variations are caused by the
decaying radiation continuum of the remnant. While  a massive
progenitor is still assumed in this scenario,  an highly
iron-enriched medium (and therefore the associated SN event) is no
more required, because of the higher densities and the greater
reprocessing efficiency in producing a line with the requested
luminosity ($L_{Fe}\approx10^{44-45} erg s^{-1}$)\citep[see][for a
comparative discussion on the reprocessor scenario]{kmr02}.

Finally, in the {\it shock heated} scenario, the gas is
collisionally ionized by shock resulting from the interaction of
GRB ejecta (either the fireball itself or the non-relativistic
ejecta of SN-like explosion associated with the GRB) with the
material pre-ejected before the GRB  by massive progenitor systems
\citep[e.g.][]{vpps_line99,b_line00,bf_line01}. The presence of an
iron recombination edge {\it in emission} is not straightforwardly
explained in this scenario, because in the aforementioned models
the plasma is expected to be near to thermal equilibrium.  However
\citet{ymm+01} argue that, under certain conditions, a plasma in a
non-equilibrium state (NEI) can also produce a narrow RRC.



\section{Star forming regions, X-ray absorption  and High-Density
Environment}

As discussed above, the presence of X-ray lines links long GRB to
massive progenitors, and therefore to star-forming sites. There is
further independent evidence supporting this connection. In the
case of NS-NS coalescence, a substantial fraction of events should
take place far from the center of the host galaxy, while the
opposite applies to massive progenitors. \citet{bkd02} have
measured the distribution of the offsets of optical afterglows
with respect to their host galaxy, finding that it is fully
consistent with that of star-forming regions, while the delayed
merging scenario can be ruled out  at the $2\times 10^{-3}$ level.

What are the implications of this scenario in terms of observable
quantities? In a typical Giant Molecular Cloud harboring star
formation in our Galaxy,  densities are $n\approx 10^2-10^5$
cm$^{-3}$, the size is of order of 10 pc and the column density is
$N_H\approx 5-10\times10^{21}$cm$^{-2}$ \citep{bwc01,wsha94}.

Such values of column density can be measured by  current X-ray
satellites in relatively bright X-ray afterglows with $z\lsim3$
(note in fact that $N_{Hobs}\approx(1+z)^{-8/3}N_{Hrest}$ and the
typical uncertainty on $N_Hobs$ is $\approx 10^{20}$cm$^{-2}$).
This is in fact the case. In the BeppoSAX sample  at least three
afterglows (GRB980703 \citep{vgo+99,dp02,s+02}; GRB010222
\citep{iz+01,dp02,s+02} and GRB990123\citep{dp02,s+02}), show
significant absorption, and two other some marginal evidence.
$N_{Hrest}$ in the range $(2-20) \nh$. In addition, Chandra (and
BeppoSAX) observations of two other afterglows give similar
results (GRB000210:$N_{Hrest}=(5\pm1)\nh$ \citep{pfg+02};
GRB000926: $N_{Hrest}=(4_{-2.5}^{+3.5}) \nh$\cite{pgg+01}).
These measurements of absorption  provide further support to the
scenario in which GRB are embedded in a star-forming GMC, in which
the   typical volume density is $n\approx10^{2-5} \dens$.

However, application of the standard fireball-jet model to
multi-wavelength data of afterglows leads to density estimates
that are typically much lower that that expected in a GMC
\citep[e.g.][]{pk01}. There are two points that need to be
stressed in this regard. First, the achromatic (i.e.
energy-independent) break observed in the light curve of some
afterglows is attributed to a collimated fireball. The break
  appears when the relativistic beaming
angle $1/\Gamma$ becomes $\approx\theta$ (e.g. Rhoads 1997, Sari
\etal 1999). The typical opening angle of the jet derived in these
models is of $\approx2-4^{\circ}$ \citep[e.g.][]{pk01,fks+01}.
 We have performed a systematic analysis of the X-ray
spectra and light curves on a sample of BeppoSAX afterglows
observed from few hours to about 2 days after the GRB \citep{s+01}
that shows that the fireball expansion {\it in the first two days}
is consistent with a spherical expansion. This result implies that
the average opening angle of the jet should be
$\theta\gsim10-20^{\circ}$. As for a probable origin of the
discrepancy we note that in  \citep[e.g.][]{pk01} the X-ray data
are considered at just one energy, overlooking the fact that the
X-ray window span almost two decades in energy, and therefore not
including the {\it X-ray spectral} information.

For example, in the case of GRB010222, where a break in the light
curves appears around 0.5 days \citep{iz+01,mpp+01}, the
predictions of the standard jet model are not consistent with the
X-ray temporal {\it and} spectral slopes \citep{iz+01}. On the
contrary, the X-ray data are well described by a fireball
undergoing a transition to a non-relativistic expansion (NRE,
\citep{lw00}). In fact, NRE can also produce an {\it achromatic
break} in the afterglow light curves at $t_{NRE} \approx 3
\frac{1+z}{2} (\frac{E_{53}}{n_6})^{1/3} {\rm days}$. This
alternative explanation has been proposed in a few other cases
(GB990123 \citep{dl99},  GRB000926 \citep{pgg+01} - but see
also\citep{hys+01} , GRB010222 \citep{iz+01,mpp+01}.  A strong
implication of this scenario is that, in those GRB's, the
environment is composed of a dense medium ($n\approx10^{4}-10^{6}
cm^{-3}$), i.e. typical of molecular clouds in star forming
regions. Interestingly, in the case of GRB010222, a strong,
constant sub-mm emission, has been attributed to enhanced
star-forming activity ($\approx 600$ \Ms yr$^{-1}$) in the galaxy
hosting this burst\citep{fbm+01}. Such high densities require a
low magnetic field, $\lsim 10^{-6}$ times the equipartition value,
to keep the synchrotron self-absorption frequency in the observed
region. Comparable values of the magnetic field have been also
derived in other cases, like GRB971214 \cite{wg99} and GRB990123
\cite{gbw+99}. However, a first attempt to fit the broad-band data
of GRB000926 with the NRE show that the radio data are not well
described by the model \citep{hys+01}.

In conclusion, at the present stage of analysis, both the NRE of a
moderately-collimated fireball in a dense medium and the highly
collimated jet scenario (in a low density medium) show some
inconsistency with the data, that appears when the complete
information from radio to X-rays is considered. More theoretical
efforts are thus needed, in particular to reconcile the fireball
model with an external medium typical of a star-forming region.

\begin{figure}
\centering
\includegraphics[width=0.5\textwidth,origin=c,angle=270]{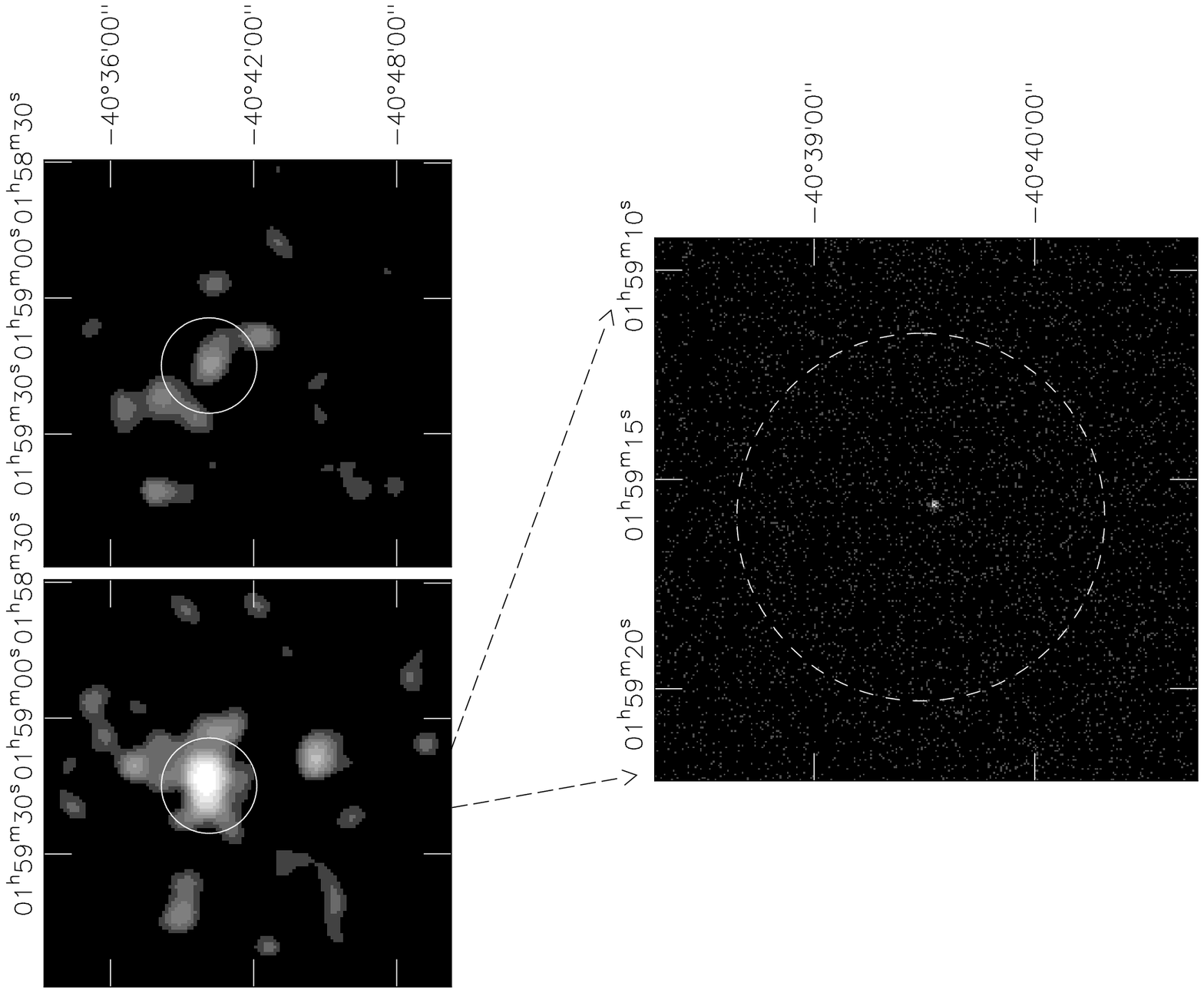}
\includegraphics[width=0.5\textwidth,clip]{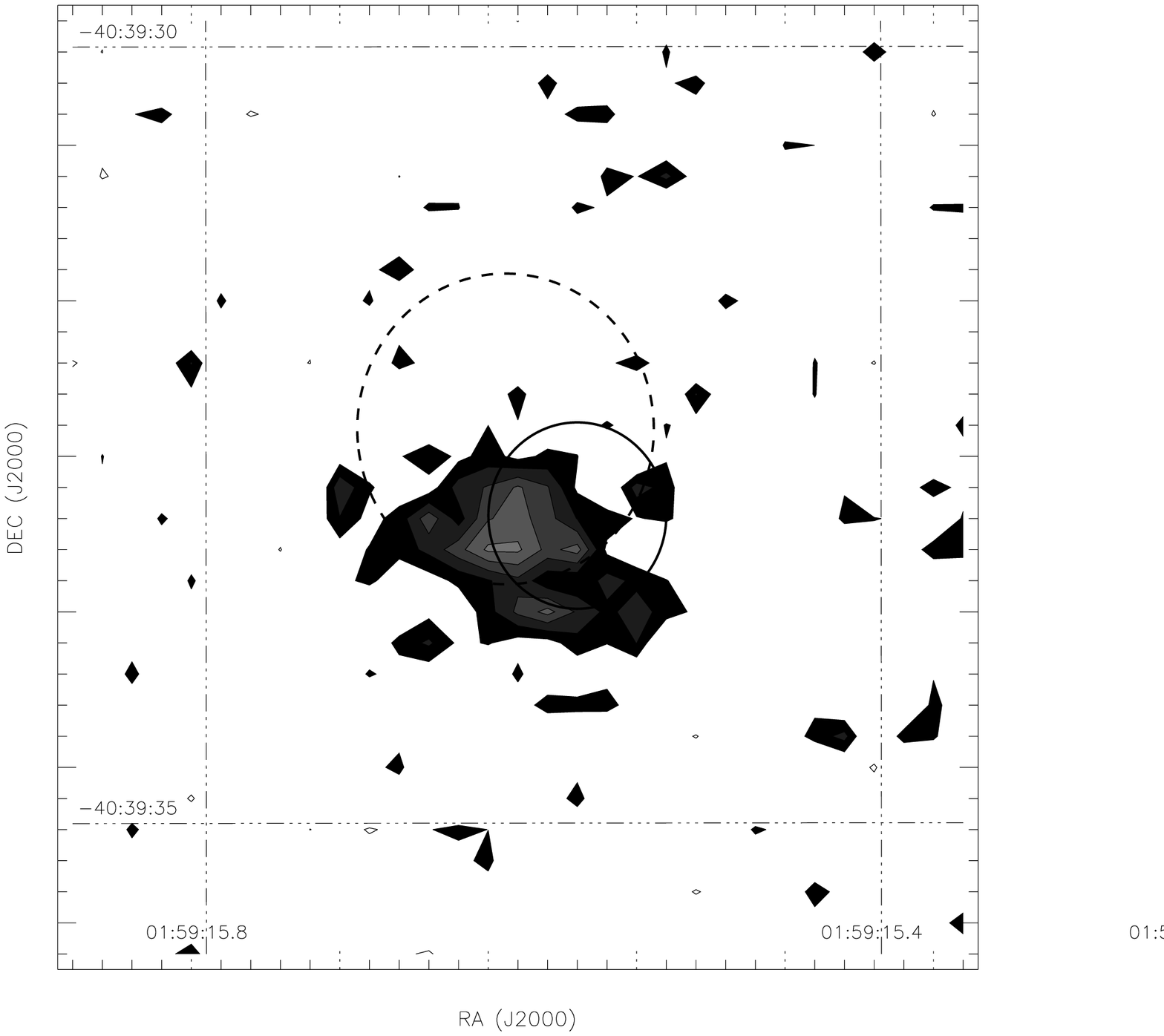}
\caption{The dark GRB000210. The two images in the left upper
panel show the decaying X-ray afterglow  BeppoSAX MECS(1.6-10 keV)
8 hrs and 30 hrs after the GRB respectively. The circle is the WFC
error box. The left lower panel is the Chandra ACIS-S image of the
afterglow 21 hrs after the GRB. The dashed line is the BeppoSAX
MECS error box. The right panel shows the optical image of the
host galaxy taken with the VLT. The circles show the 90\% error
circles of the Chandra (continuous line) and radio (dashed line)
afterglows. The redshift of the galaxy is z=0.846}
\label{fig:gb000210}
\end{figure}

\section{Recent findings and remaining mysteries}

\subsection{GHOST and X-ray flashes }

It is observationally well-established that about half of
accurately localized  gamma-ray bursts (GRBs) do not produce a
detectable optical afterglow \citep{fkw+00,fjg+01}, while most of
them ($\approx90\%$) have an X-ray afterglow \citep{piro01}.
Statistical studies have shown that the optical searches of these
events, known variously as ``dark GRBs'', ``failed optical
afterglows'' (FOA), or ``gamma-ray bursts hiding an optical
source-transient'' (GHOST), have been carried out to
  magnitude limits fainter on average than the  known sample of
optical afterglows \citep{lcg01,ry01}. Thus dark bursts appear to
constitute a distinct class of events and  are not the result of
an inadequate search, but it is unclear whether  this
observational property derives from a  single origin  or it is a
combination of different causes.

If the progenitors of long-duration GRBs are massive stars
\citep{pac98b}, as current evidence suggests ({\em e.g.,}
\citep{bkd+99,pgg+00}), extinction of optical flux by dusty
star-forming regions is likely to occur for a substantial fraction
of events ({\it the obscuration scenario}\citep[e.g.][]{ry01})).
Another possibility is that dark GRBs are located at redshift
$z\gsim$5, with the optical flux being absorbed by the intervening
Ly$\alpha$ forest clouds ({\it the
  intergalactic scenario} \citep[e.g.][]{fru99}).

Dark bursts which can be localized to arcsecond accuracy, through
a detection of either   their X-ray or radio afterglow, are of
particular interest.  The first example is GRB\,970828 for which
prompt, deep searches down to R$\sim$24.5 failed to detect an
optical afterglow \citep{odk+97, ggv+98d} despite it was localized
within a region of only 10 arcsec radius by the ROSAT satellite
\citep{gse+97}. \citet{dfk+01} recently showed how the detection
of a short-lived radio transient for GRB\, 970828 allowed them to
identify the probable host galaxy and to infer its properties
(redshift, luminosity and morphology). In addition, they used
estimates of the column density of absorbing gas from X-ray data,
and lower limits on the rest frame extinction (A$_V>3.8$) to
quantify the amount of obscuration towards the GRB. The other best
studied example is GRB000210 \citep{pfg+02}. This burst had the
highest $\gamma$-ray peak flux of any  event localized by {\em
BeppoSAX} as yet but it did not have a detected optical afterglow,
despite prompt and deep searches down to $R_{lim}\approx 23.5$.
{\em Chandra} observations allowed  to localize the X-ray
afterglow of GRB000210 to within $\approx 1$arcsec and a radio
transient was detected with the VLA. The precise X-ray and radio
positions allowed to identify the likely host galaxy of this
burst, and to measure its redshift, $z=0.846$. The probability
that this galaxy is a field object is $\approx 10^{-2}$, but the
chance that this happens in both the cases of GRB970828 \&
GRB000210 is negligible.

The X-ray spectrum of the afterglow of GRB000210 shows significant
absorption in excess of the Galactic one corresponding, at the
redshift of the galaxy, to $N_H=(5\pm1)\times 10^{21}$ cm$^{-2}$.
The amount of dust needed to absorb the optical flux of this
object is consistent with the above HI column density, given a
dust-to-gas ratio similar to that of our Galaxy. If the absorption
takes place in a GMC, a substantial fraction of the X-ray gas
should be heavily ionized by the hard X-ray photons emitted by the
GRB and its afterglow. This effect should change the absorption
profile, in particular at lower energies, where the lighter
elements are heavily ionized, and therefore become more
transparent to the radiation. Nonetheless, the X-ray absorption
observed in GRB000210 is consistent with a cold medium, an
evidence that can be reconciled with the GMC scenario is the
medium is condensed in high dense clouds ($n\gsim10^9$ cm$^{-3}$
\citep{pfg+02}.

Given the extreme luminosity of GRBs and their probable
association with massive stars, it is expected that some fraction
of events will be located beyond $z>5$ \citep{lr00}. These would
be probably classified as dark bursts because the UV light, which
is strongly attenuated by absorption in the Ly$_\alpha$ forest, is
redshifted into the optical band.  We note that the four redshifts
determined or suggested so far for dark GRBs ($z=0.96$, GRB970828,
\citep{dfk+01};$z=1.3$, GRB990506, \citep{tbf+00, bkd02};
$z\approx0.47$, GRB000214, \citep{apv+00}, GRB000210
\citep{pfg+02}) are in the range of those measured for most bright
optical afterglows, but whether this applies to the majority of
these events is still to be assessed.

Particularly interesting in this respect is the case of the
so-called X-ray flashes or  X-ray rich GRBs discovered by {\em
BeppoSAX}\citep{hzkw01,h+02}. No optical counterpart has been
identified as yet in any of these bursts. The {\it intergalactic
scenario} would naturally explain both the absence of optical
afterglows and the high-energy spectrum, because the peak of the
gamma-ray spectrum would be redshifted into the X-ray band.
Alternatively, the paucity of gamma-ray emission can be explained
by a "dirty" fireball \citep{dermer_dirty99}. In such a case the
fireball will achieve a Lorentz factor much lower than previously
considered, i.e. not high enough to boost the photons in the
$\gamma$-ray range.

\subsection{Short GRB's}

The distribution of GRB duration appears to be bimodal, with about
$30\%$ of events lasting less than 1 sec \citep{kmf+93}. It is
still unclear whether these events are intrinsically diverse from
long bursts, i.e. if they are produced by different progenitors.
So far, very little is known about these events, due to the lack
of a counterpart. While the GRBM on board of BeppoSAX is detecting
those events, no one has so far been observed in the X-ray range
(and therefore precisely localized) by the WFC \cite{g_short+01}.
The probability that this is a chance fluctuation is getting
interestingly low, suggesting a  depleted X-ray emission compared
to the class of long events \cite{g+02}.  If that is the case, the
localization of afterglows of these events should rely only on
experiments with good position accuracy in the $\gamma$-ray range,
like the currently operating IPN or, in the future,  SWIFT. If,
instead, these events have an X-ray emission similar to long ones,
we may hope to localize a few of them with BeppoSAX and HETE2.

\section{The future of X-ray spectroscopy}

X-ray spectroscopy of GRB is coming of age. Emission and
absorption features as well as the properties of X-ray absorption
are providing key information on the close environment of long
GRB's, suggesting massive progenitors and a connection between
these events and star-forming regions. The possibility of
measuring the redshift directly from X-rays is of particular value
for those classes of objects, like the majority of dark GRB, X-ray
flashes and short GRB's, that still lack optical counterparts. The
radiation intensity of GRB's  is so high that they can be
detectable out to much larger distances than those of the most
luminous quasars or galaxies observed so far, and it is likely
that  high-z GRB are actually the constituents of one of those
mysterious classes of events.

 We should stress that, with  present
X-ray facilities (BeppoSAX, XTE, HETE2, Chandra, Newton), the
progress in this field will follow "quantum" jumps. In fact, to
get good-quality X-ray spectra, one should catch bright
afterglows, but these should be not too many, because of the
combination of ingredients needed (number of precise and fast
locations, reaction time, number of TOO observations allocated to
GRB programs). On the positive side, one can then hope that the
next important discovery just lies behind the corner.

 In the near future, we expect a further advancement in this
area, when high quality X-ray CCD spectra of afterglows ($E/\Delta
E\approx50$) will be routinely available with SWIFT.
 Looking ahead
in the future, high resolution X-ray spectroscopy ($E/\Delta
E\gsim1000$, like that provided by X-ray microcalorimeters) should
open a new area of exploration (\cite{p_imboss01,p_xeus}), that
would bring us closer to the central engine of GRB, its
environment and far in the Early Universe by using GRB as beacons
to probe star and galaxy formation. We can foresee the possibility
of resolving in detail line profiles, deriving information on the
kinematics of the ejecta, or looking for narrow and faint emission
lines, imprinted on the spectra by the low-velocity medium
embedding the GRB. The measurement of the absorption edges
produced by the ISM of the {\it host galaxy} will provide
information on the chemical composition of galaxies in the Early
Universe, thus opening the possibility to trace the metallicity
history of the gas in the Universe and to probe the formation of
the first stars \citep{p_xeus,f+00,lr00}.

\begin{theacknowledgments}
BeppoSAX is a program of the Italian space agency (ASI) with
participation of the Dutch space agency (NIVR)
\end{theacknowledgments}



\begin{thebibliography}{63}
\expandafter\ifx\csname natexlab\endcsname\relax\def\natexlab#1{#1}\fi
\providecommand{\enquote}[1]{``#1''}
\expandafter\ifx\csname url\endcsname\relax
  \def\url#1{\texttt{#1}}\fi
\expandafter\ifx\csname urlprefix\endcsname\relax\def\urlprefix{URL }\fi

\bibitem[Woosley(2001)]{w01}
Woosley, S.~E.,  in \emph{GRBs in the
  Afterglow Era}, edited by E.~Costa, F.~Frontera, and J.~Hjorth, ESO-Springer,
  2001, pp. 258--262.

\bibitem[Reichart and Yost(2001)]{ry01}
Reichart, D.~E., and Yost, S.~A., 
  (2001), apJ, in press; astro-ph/0107545.

\bibitem[Dermer et~al.(1999)]{dermer_dirty99}
Dermer, C.~D., Chiang, J., and Boettcher, M., \emph{ApJ}, \textbf{513},
  656--668 (1999).

\bibitem[Piro et~al.(1998)]{paa+98}
Piro, L., et~al., \emph{A\&A}, \textbf{331}, L41--L44 (1998).

\bibitem[{Yoshida} et~al.(2001)]{yym+01}
{Yoshida}, A., {Yonetoku}, N.~M., {Murakami}, T., {Otani}, C., {Kawai}, N.,
  {Ueda}, Y., {Shibata}, R., and {Uno}, S., \emph{ApJ}, \textbf{557}, L27--L30
  (2001).

\bibitem[{Antonelli} et~al.(2000)]{apv+00}
{Antonelli}, L.~A., {Piro}, L., {Vietri}, \etal \emph{ApJ}, \textbf{545}, L39--L42 (2000).

\bibitem[{Piro} et~al.(2000)]{pgg+00}
{Piro}, L., {Garmire}, G., {Garcia}, \etal \emph{Science}, \textbf{290}, 955--958 (2000).

\bibitem[{Amati} et~al.(2000)]{afv+00}
{Amati}, L., {Frontera}, F., {Vietri}, \etal \emph{Science}, \textbf{290}, 953--955 (2000).

\bibitem[Piro(1993)]{piro93}
Piro, L.,  in \emph{UV and X-ray spectroscopy of Laboratory and
  Astrophysical Plasmas}, edited by S.~K. E.~Silver, Cambridge Uni. Press,
  1993, p. 448.

\bibitem[Andersen et~al.(2002)]{andersen_990705}
Andersen, M., et~al. (2002), in preparation.

\bibitem[Lazzati et~al.(1999)]{lcg_line99}
Lazzati, D., Campana, S., and Ghisellini, G., \emph{MNRAS}, \textbf{304}, L31
  (1999).

\bibitem[{Weth} et~al.(2000)]{weth+00}
{Weth}, C., et~al., \emph{ApJ}, \textbf{534}, 581 (2000).

\bibitem[{Perna} and {Loeb}(1998)]{pl98}
{Perna}, R., and {Loeb}, A., \emph{ApJ}, \textbf{501}, 467-- (1998).

\bibitem[{Boettcher} et~al.(1999)]{bdcl99}
{Boettcher}, M., {Dermer}, C.~D., {Crider}, A.~W., and {Liang}, E. .~P.,
  \emph{A\&A}, \textbf{343}, 111--119 (1999).

\bibitem[{Paerels} et~al.(2000)]{pkhl00}
{Paerels}, F., {Kuulkers}, E., {Heise}, J., and {Liedahl}, D.~A., \emph{ApJ},
  \textbf{535}, L25--L28 (2000).

\bibitem[{Vietri} and {Stella}(1999)]{vs_supra99}
{Vietri}, M., and {Stella}, L., \emph{ApJ}, \textbf{527}, L43--L46 (1999).

\bibitem[{Rees} and {M{\'e}sz{\' a}ros}(2000)]{rm_line00}
{Rees}, M.~J., and {M{\'e}sz{\' a}ros}, P., \emph{ApJ}, \textbf{545}, L73--L75
  (2000).

\bibitem[{M{\'e}sz{\' a}ros} and {Rees}(2001)]{mr_line01}
{M{\'e}sz{\' a}ros}, P., and {Rees}, M.~J., \emph{ApJ}, \textbf{556}, L37--L40
  (2001).

\bibitem[Kallman et~al.(2002)]{kmr02}
Kallman, T.~R., M{\'e}sz{\' a}ros, P., and Rees, M.~J., Iron k lines from grb
  (2002), astro-ph/00110654.

\bibitem[{Vietri} et~al.(1999)]{vpps_line99}
{Vietri}, M., {Perola}, G.~C., {Piro}, L., and {Stella}, L., \emph{MNRAS},
  \textbf{308}, L29 (1999).

\bibitem[{Boettcher}(2000)]{b_line00}
{Boettcher}, M., \emph{ApJ}, \textbf{539}, 102--110 (2000).

\bibitem[{Boettcher} and {Fryer}(2001)]{bf_line01}
{Boettcher}, M., and {Fryer}, C.~L., \emph{ApJ}, \textbf{547}, 338--344 (2001).

\bibitem[{Yonetoku} et~al.(2001)]{ymm+01}
{Yonetoku}, N.~M., {Murakami}, T., {Masai}, K., {Yoshida}, A., {Kawai}, N., and
  {Namiki}, M., \emph{ApJ}, \textbf{557}, L23--L26 (2001).

\bibitem[Bloom et~al.(2002)]{bkd02}
Bloom, J.~S., Kulkarni, S.~R., and Djorgovski, S.~G.(2002), aJ, in press; astro-ph/0010176.

\bibitem[{Balsara} et~al.(2001)]{bwc01}
{Balsara}, D., {Ward-Thompson}, D., and {Crutcher}, R.~M., \emph{MNRAS},
  \textbf{327}, 715--720 (2001).

\bibitem[{Ward-Thompson} et~al.(1994)]{wsha94}
{Ward-Thompson}, D., {Scott}, P.~F., {Hills}, R.~E., and {Andre}, P.,
  \emph{MNRAS}, \textbf{268}, 276--290 (1994).

\bibitem[{Vreeswijk} et~al.(1999)]{vgo+99}
{Vreeswijk}, P.~M., \etal \emph{ApJ}, \textbf{523}, 171--176 (1999).

\bibitem[Pasquale et~al.(2002)]{dp02}
Pasquale, M.~D., et~al. (2002), in preparation.

\bibitem[Stratta et~al.(2002)]{s+02}
Stratta, G., et~al. (2002), in preparation.

\bibitem[{in' t Zand} et~al.(2001)]{iz+01}
{in' t Zand}, J. . J.~M., {Kuiper}, L., {Amati}\etal \emph{ApJ}, \textbf{559},
  710--715 (2001).

\bibitem[Piro et~al.(2002)]{pfg+02}
Piro, L., Frail, D., Gorosabel, J., \etal (2002), apJ, submitted;
  astro/ph-0201282.

\bibitem[{Piro} et~al.(2001)]{pgg+01}
{Piro}, L., {Garmire}, G., {Garcia}, \etal \emph{ApJ}, \textbf{558}, 442--447 (2001).

\bibitem[{Panaitescu} and {Kumar}(2001)]{pk01}
{Panaitescu}, A., and {Kumar}, P., \emph{ApJ}, \textbf{554}, 667--677 (2001).

\bibitem[Frail et~al.(2001{\natexlab{a}})]{fks+01}
Frail, D.~A., et~al., \emph{ApJ}, \textbf{562}, L55-- (2001{\natexlab{a}}).

\bibitem[Stratta et~al.(2001)]{s+01}
Stratta, G., Piro, L., Soffitta, P., et~al.,  in \emph{GRBs in the
  Afterglow Era}, edited by E.~Costa, F.~Frontera, and J.~Hjorth, ESO-Springer,
  2001, pp. 118--120.

\bibitem[Masetti et~al.(2001)]{mpp+01}
Masetti, N., et~al., \emph{A\&A}, \textbf{374}, 382 (2001).

\bibitem[{Livio} and {Waxman}(2000)]{lw00}
{Livio}, M., and {Waxman}, E., \emph{ApJ}, \textbf{538}, 187--191 (2000).

\bibitem[{Dai} and {Lu}(1999)]{dl99}
{Dai}, Z.~G., and {Lu}, T., \emph{ApJ}, \textbf{519}, L155--L158 (1999).

\bibitem[Harrison et~al.(2001)]{hys+01}
Harrison, F.~A.,\etal \textbf{559}, 523 (2001).

\bibitem[Frail et~al.(2001{\natexlab{b}})]{fbm+01}
Frail, D.~A., \etal (2001{\natexlab{b}}), submitted to
  ApJ; astro-ph/0108436.

\bibitem[{Wijers} and {Galama}(1999)]{wg99}
{Wijers}, R. A. M.~J., and {Galama}, T.~J., \emph{ApJ}, \textbf{523}, 177--186
  (1999).

\bibitem[{Galama} et~al.(1999)]{gbw+99}
{Galama}, T.~J., \etal,
  \emph{Nature}, \textbf{398}, 394--399 (1999).

\bibitem[{Frail} et~al.(2000)]{fkw+00}
{Frail}, D.~A., {Kulkarni}, S.~R., {Wieringa}, M.~H., {Taylor}, G.~B.,
  {Moriarty-Schieven}, G.~H., {Shepherd}, D.~S., {Wark}, R.~M., {Subrahmanyan},
  R., {McConnell}, D., and {Cunningham}, S.~J., , in \emph{AIP Conf. Proc. 526: Gamma-ray Bursts, 5th
  Huntsville Symposium}, 2000, pp. 298--302.

\bibitem[{Fynbo} et~al.(2001)]{fjg+01}
{Fynbo}, J.~U., \etal
  \emph{A\&A}, \textbf{369}, 373--379 (2001).

\bibitem[Piro(2001)]{piro01}
Piro, L.,  in \emph{GRBs in the
  Afterglow Era}, edited by E.~Costa, F.~Frontera, and J.~Hjorth, ESO-Springer,
  2001, pp. 97--105.

\bibitem[Lazzati et~al.(2002)]{lcg01}
Lazzati, D., Covino, S., and Ghisellini, G.,  (2002), mNRAS, in press,
  astro-ph/0011443.

\bibitem[Paczy\'nski(1998)]{pac98b}
Paczy\'nski, B., \emph{ApJ}, \textbf{494}, L45--L48 (1998).

\bibitem[Bloom et~al.(1999)]{bkd+99}
Bloom, J.~S., et~al., \emph{Nature}, \textbf{401}, 453--456 (1999).

\bibitem[Fruchter(1999)]{fru99}
Fruchter, A.~S., \emph{ApJ}, \textbf{512}, L1--L4 (1999).

\bibitem[{Odewahn} et~al.(1997)]{odk+97}
{Odewahn}, S.~C., \etal \emph{IAU Circ}, \textbf{6735} (1997).

\bibitem[{Groot} et~al.(1998)]{ggv+98d}
{Groot}, P.~J.\etal \emph{ApJ}, \textbf{493},
  L27--+ (1998).

\bibitem[{Greiner} et~al.(1997)]{gse+97}
{Greiner}, J., {Schwarz}, R., {Englhauser}, J., {Groot}, P.~J., and {Galama},
  T.~J., \emph{IAU Circ}, \textbf{6757} (1997).

\bibitem[Djorgovski et~al.(2001)]{dfk+01}
Djorgovski, S.~G., , Frail, D.~A., Kulkarni, S.~R., Bloom, J.~S., Odewahn,
  S.~C., and Dierks, A., \emph{ApJ}, \textbf{562}, 654 (2001).

\bibitem[{Lamb} and {Reichart}(2000)]{lr00}
{Lamb}, D.~Q., and {Reichart}, D.~E., \emph{ApJ}, \textbf{536}, 1--18 (2000).

\bibitem[{Taylor} et~al.(2000)]{tbf+00}
{Taylor}, G.~B., {Bloom}, J.~S., {Frail}, D.~A., {Kulkarni}, S.~R.,
  {Djorgovski}, S.~G., and {Jacoby}, B.~A., \emph{ApJ}, \textbf{537}, L17--L21
  (2000).

\bibitem[Heise et~al.(2001)]{hzkw01}
Heise, J., in~'t Zand, J., Kippen, M., and Woods, P.,  in \emph{GRBs in the Afterglow Era}, edited
  by E.~Costa, F.~Frontera, and J.~Hjorth, ESO-Springer, 2001, pp. 16--21.

\bibitem[Heise(2002)]{h+02}
Heise, J. (2002), this conference.

\bibitem[Kouveliotou et~al.(1993)]{kmf+93}
Kouveliotou, C., Meegan, C.~A., Fishman, G.~J., Bhat, N.~P., Briggs, M.~S.,
  Koshut, T.~M., Paciesas, W.~S., and Pendleton, G.~N., \emph{ApJ},
  \textbf{413}, 101--104 (1993).

\bibitem[Gandolfi et~al.(2000)]{g_short+01}
Gandolfi, G., et~al.,  in \emph{AIP Conf. Proc. 526: Gamma-ray
  Bursts, 5th Huntsville Symposium}, edited by M.~Kippen, R.~Mallozzi, and
  G.~J. Fishman, 2000, pp. 23--.

\bibitem[Gandolfi et~al.(2002)]{g+02}
Gandolfi, G., et~al. (2002), this conference.

\bibitem[Piro et~al.(2001)]{p_imboss01}
Piro, L., et~al.,in \emph{GRBs in the Afterglow Era}, edited by
  E.~Costa, F.~Frontera, and J.~Hjorth, ESO-Springer, 2001, p. 415.

\bibitem[Piro(1999)]{p_xeus}
Piro, L. (1999), probing the Early Universe with GRB and XEUS at
  www.ias.rm.cnr.it/sax/xeus.html.

\bibitem[{Fiore} et~al.(2000)]{f+00}
{Fiore}, F., et~al., \emph{ApJ}, \textbf{544}, L7-- (2000).

\end{thebibliography}

\end{document}